\documentclass[12pt]{article}
\usepackage{epsfig}\parskip 5pt plus 1pt
\usepackage{eurosym}
\usepackage{ulem}
\usepackage{color}
\usepackage{setspace}
\usepackage{amssymb}
\newcommand{\be}{\begin{equation}}
\newcommand{\ee}{\end{equation}}
\newcommand{\bea}{\begin{eqnarray}}
\newcommand{\eea}{\end{eqnarray}}

\def\simge{\mathrel{%
   \rlap{\raise 0.511ex \hbox{$>$}}{\lower 0.511ex \hbox{$\sim$}}}}
\def\simle{\mathrel{
   \rlap{\raise 0.511ex \hbox{$<$}}{\lower 0.511ex \hbox{$\sim$}}}}

\newcommand{\ENUBET}{\textsc{Enubet~}}
\newcommand{\ENUBETnospace}{\textsc{Enubet}}

\newcommand{\ENUBETcomma}{\textsc{Enubet},}

\begin{document}
\thispagestyle{empty}
\begin{center}
  {\large{\bf Input document for the European Particle Physics Strategy} }\\
\vspace*{0.5cm}

{\Large{\bf A high precision neutrino beam for a new generation of short baseline experiments} }\\
\end{center}

\noindent
F.~Acerbi$^{a}$,
G.~Ballerini$^{b,o}$,
S.~Bolognesi$^{t}$,
M.~Bonesini$^{b}$,
C.~Brizzolari$^{b,o}$,
G.~Brunetti~$^j$
S.~Carturan~$^{j,k}$,
M.G.~Catanesi$^l$,
S.~Cecchini$^c$,
F.~Cindolo$^c$,
G.~Collazuol$^{j,k}$,
E.~Conti$^j$,
F.~Dal Corso$^j$,
G.~De Rosa$^{p,q}$,
F.~Di Lodovico$^{v}$,
C.~Delogu$^{b,h}$,
A.~Falcone$^{j,k}$,
A.~Gola$^a$,
R.A.~Intonti$^l$,
C.~Jollet$^d$,
B.~Klicek$^s$,
Y.~Kudenko$^r$,
M.~Laveder$^j$,
A.~Longhin$^{j,k(*)}$,
L.~Ludovici$^f$,
L.~Magaletti$^l$,
G.~Mandrioli$^c$,
A.~Margotti$^c$,
V.~Mascagna$^{b,o}$,
N.~Mauri$^c$,
A.~Meregaglia$^d$,
M.~Mezzetto$^j$,
M.~Nessi$^m$,
A.~Paoloni$^e$,
M.~Pari$^{j,k,m}$,
E.~Parozzi$^{b,h}$,
L.~Pasqualini$^{c,g}$,
G.~Paternoster$^a$,
L.~Patrizii$^c$,
C.~Piemonte$^a$,
M.~Pozzato$^c$,
F.~Pupilli$^j$,
M.~Prest$^{o,b}$,
E.~Radicioni$^l$,
C.~Riccio$^{p,q}$,
A.C.~Ruggeri$^{p,q}$,
F.~Sanchez Nieto$^u$,
G.~Sirri$^c$,
M.~Soldani$^{o,b}$,
M.~Stipcevic$^s$,
M.~Tenti$^{b,h}$,
F.~Terranova$^{b,h}$,
M.~Torti$^{b,h}$,
E.~Vallazza$^i$,
M.~Vesco$^k$,
L.~Votano$^e$.

\noindent
$^a$ Fondazione Bruno Kessler (FBK) and INFN TIFPA, Trento, Italy.\\
$^b$ INFN, Sezione di Milano-Bicocca, Piazza della Scienza 3, Milano, Italy.\\
$^c$ INFN, Sezione di Bologna, viale Berti-Pichat 6/2, Bologna, Italy.\\
$^d$ CENBG, Universit\`e de Bordeaux, CNRS/IN2P3, 33175 Gradignan, France.\\
$^e$ INFN, Laboratori Nazionali di Frascati, via Fermi 40, Frascati (Rome), Italy.\\
$^f$ INFN, Sezione di Roma 1, piazzale A. Moro 2, Rome, Italy.\\
$^g$ Phys. Dep. Universit\`a di Bologna, viale Berti-Pichat 6/2, Bologna, Italy.\\
$^h$ Phys. Dep. Universit\`a di Milano-Bicocca, Piazza della scienza 3, Milano, Italy.\\
$^i$ INFN Sezione di Trieste, via Valerio, 2 - Trieste, Italy.\\
$^j$ INFN Sezione di Padova, via Marzolo, 8 - Padova, Italy.\\
$^k$ Phys. Dep. Universit\`a di Padova, via Marzolo, 8 - Padova, Italy.\\
$^l$ INFN Sezione di Bari, via Amendola, 173 - Bari, Italy.\\
$^m$ CERN, Geneva, Switzerland.\\
$^n$ Phys. Dep. Universit\`a La Sapienza, piazzale A. Moro 2, Rome, Italy.\\
$^o$ DISAT, Universit\`a degli Studi dell'Insubria, via Valeggio 11, Como, Italy.\\
$^p$ INFN, Sezione di Napoli, Via Cintia, Napoli, Italy.\\
$^q$ Phys. Dep., Universit\`a ``Federico II'' di Napoli, Napoli, Italy.\\
$^r$ Institute of Nuclear Research of the Russian Academy of Science, Moscow, Russia.\\
$^s$ 
CEMS, Rudjer Boskovic Institute, HR-10000 Zagreb, Croatia.\\
$^t$ Centre CEA de Saclay, Gif-sur-Yvette 91191 cedex, France.\\
$^u$ Universit\`e de Geneve, 24, Quai Ernest-Ansermet, 1211 Geneva 4, Switzerland.\\
$^v$ Queen Mary University of London, School of Physics and Astronomy, London, UK.\\ 
$^{(*)}$ Contact person. A. Longhin (andrea.longhin@pd.infn.it).
\newpage
\begin{abstract}
\noindent
The current generation of short baseline neutrino experiments is
approaching intrinsic source limitations in the knowledge of flux,
initial neutrino energy and flavor. A dedicated facility based on
conventional accelerator techniques and existing infrastructures
designed to overcome these impediments would have a remarkable impact
on the entire field of neutrino oscillation physics. It would improve
by about one order of magnitude the precision on $\nu_\mu$ and $\nu_e$
cross sections, enable the study of electroweak nuclear physics at the
GeV scale with unprecedented resolution and advance searches for
physics beyond the three-neutrino paradigm. In turn, these results
would enhance the physics reach of the next generation long baseline
experiments (DUNE and Hyper-Kamiokande) on CP violation and their
sensitivity to new physics. In this document, we present the physics
case and technology challenge of high precision neutrino beams based
on the results achieved by the \ENUBET Collaboration in 2016-2018.  We
also set the R\&D milestones to enable the construction and running of
this new generation of experiments well before the start of the DUNE
and Hyper-Kamiokande data taking.  We discuss the implementation of
this new facility at three different level of complexity: $\nu_\mu$
narrow band beams, $\nu_e$ monitored beams and tagged neutrino
beams. We also consider a site specific implementation based on the
CERN-SPS proton driver providing a fully controlled neutrino source to
the ProtoDUNE detectors at CERN.
\end{abstract}
 
\section{Introduction}
\label{sec:introduction}

Over the last 50 years, accelerator neutrino beams~\cite{Kopp:2006ky}
have been developed toward higher and higher intensities but
uncertainties in the flux, flavor composition and initial neutrino
energy are still very large.  Thanks to the enormous progress in
neutrino scattering
experiments~\cite{Katori:2016yel,Alvarez-Ruso:2017oui}, the
measurements of neutrino cross sections are now limited by the
knowledge of the initial fluxes, since the yield of $\nu_\mu$ is not
measured in a direct manner but relies on extrapolation from
hadro-production data and a detailed simulation of the neutrino
beamline. This limitation bounds the precision that can be reached in
the measurement of the absolute cross sections to ${\cal O}
(5-10\%)$. In addition, current experiments reconstruct the neutrino
energy from final state particles. As a consequence, the reconstructed
energy and, in turn, the measurement of the cross section is affected
by model dependencies. Finally, pion-based sources mainly produce $\nu_\mu$ while most of the
next generation oscillation experiments will rely on the appearance of
$\nu_e$ at the far detector. A direct measurement of the $\nu_e$ cross
sections is, hence, of great value for the current (T2K, NO$\nu$A) and
forthcoming (DUNE, Hyper-Kamiokande) long-baseline
experiments~\cite{Ankowski:2016jdd}.

The aim of this document is to:

\begin{itemize}
\item present the physics case of a dedicated narrow-band beam, which
  addresses precision neutrino physics at short baselines;
\item foster a vigorous R\&D programme to enable the construction and
  running of a new generation of cross section experiments well before
  the start of the DUNE and Hyper-Kamiokande data taking;
\item pave the road for a CERN-based facility dedicated to high
  precision neutrino scattering physics. This facility will run in
  parallel with DUNE and Hyper-Kamiokande. It will thus enhance the
  physics reach of long-baseline experiments (in particular CP reach
  and non-standard neutrino oscillation physics) reducing in a
  substantial manner their leading systematic contributions.
\end{itemize}

We will discuss three implementations of high precision conventional
(i.e. pion/kaon based) neutrino beams, at different level of
complexity and cost. Non conventional facilities based on muon storage
and decay are beyond the scope and timeline considered in this
document and we refer to~\cite{townmeeting,Adey:2015iha} for
additional insights on these technologies.

High precision cross section physics motivated the development of
``monitored neutrino beams''~\cite{Longhin:2014yta} and, in turn, the
\ENUBET proposal~\cite{enubet_eoi,enubet_loi_2018}: a facility where
the only source of electron neutrino is the three body semileptonic
decay of kaons: $K^+ \rightarrow \pi^0 e^+ \nu_e $
($K_{e3}$). Most of the results presented in this document has been
achieved by the \ENUBET Collaboration in 2016-2018.
The ERC \ENUBET (``Enhanced NeUtrino BEams from kaon Tagging'')
project~\cite{enubet_erc} is aimed at building a detector that
identifies positrons in $K_{e3}$ decays while operating in the harsh
environment of a conventional neutrino beam decay tunnel. The project
addresses all accelerator challenges of monitored neutrino beams: the
proton extraction scheme, focusing and transfer line, instrumentation
of the decay tunnel and the assessment of the physics performance.
This technique is currently the most promising method to fulfill
simultaneously all the requirements for high precision $\nu_\mu$ and
$\nu_e$ cross section measurements.

The \ENUBET activities are embedded in a larger community effort to
improve our knowledge of neutrino properties and perform ancillary
measurements for the next generation long baseline experiments. This
framework is detailed in a dedicated input document for the European
Strategy~\cite{townmeeting}.

\section{The physics of high precision neutrino beams}

High precision neutrino beams are facilities that provide a control of
the neutrino flux at source with 1\% level precision and the beam
energy spread constrains the initial neutrino energy within $\sim$10\%.

\subsection{GeV scale interactions of neutrinos with matter}
\begin{figure}
  \centering
  \includegraphics[width=0.55\textwidth]{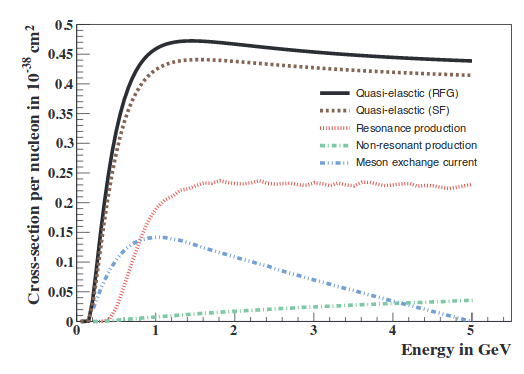}%
  \includegraphics[width=0.45\textwidth]{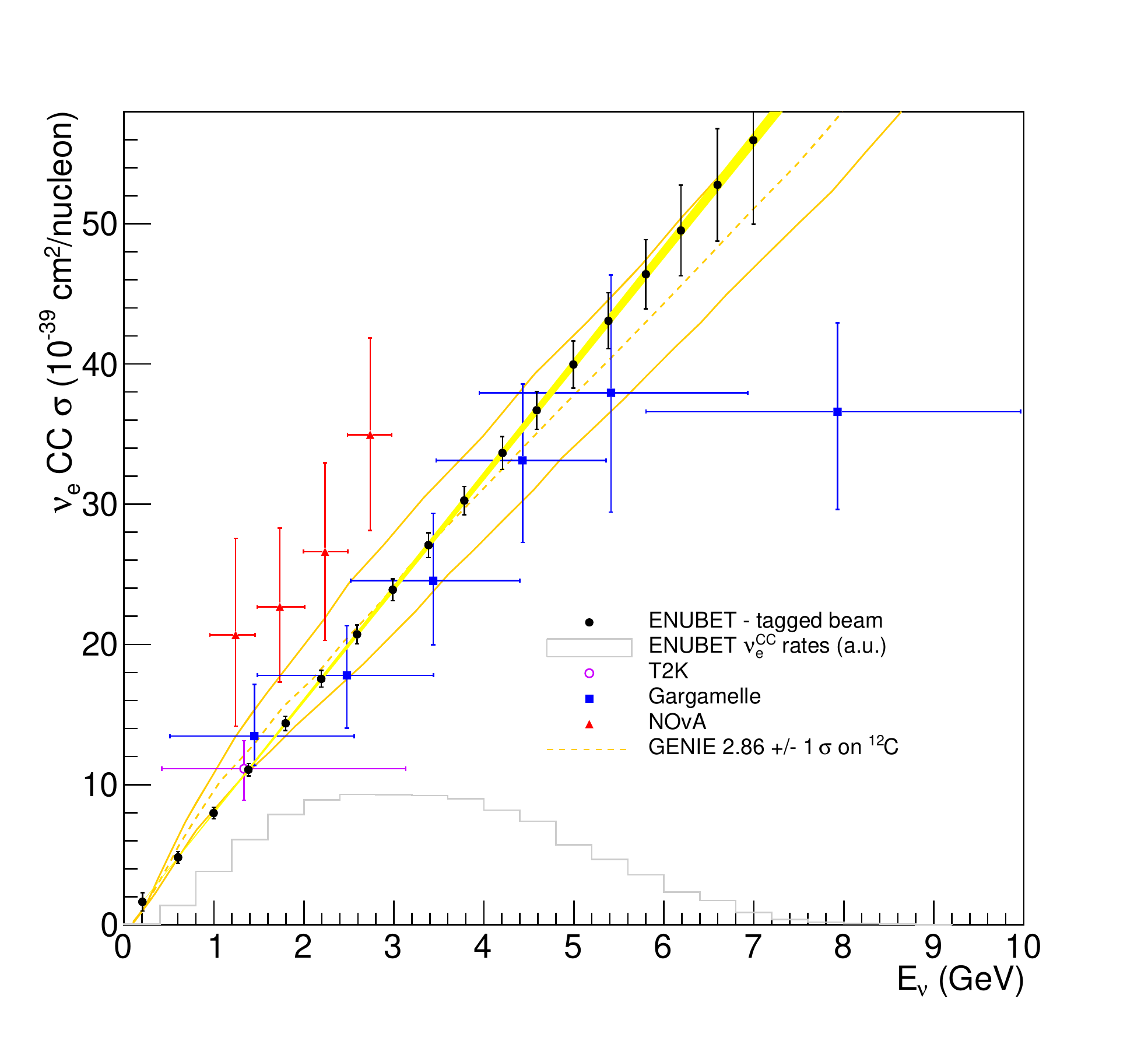}
  \caption{Left: neutrino interaction processes at the GeV scale (from
    \cite{Ankowski:2016jdd}).
  Right: present status of the electron neutrino cross section measurements (Gargamelle, NO$\nu$A, T2K), theory expectation (GENIE) and projected measurements from ENUBET in one year of data taking with Protodune-SP.}
  \label{fig:nue}
\end{figure}

When coupled to fine grained neutrino detectors, this facility could
unravel the complexity of neutrino interactions at the GeV scale~\cite{Alvarez-Ruso:2017oui}:
quasi-elastic interactions, resonance production, meson exchange
(multi-nucleon) and, at higher energies, the onset of deep inelastic
scattering off quarks (see Fig.~\ref{fig:nue} left, from
\cite{Ankowski:2016jdd}). In \ENUBET a superior level of control is
achieved reducing the overall normalization uncertainty in the flux
and introducing the constraint on the incoming energy through the
narrow-band off-axis  technique (NBOA - see
Sec.~\ref{sec:numu}).

This is particularly remarkable for the study of electron neutrino
cross sections.  Fig.~\ref{fig:nue} (right) shows the present
measurements of the $\nu_e^{CC}$ cross section. Conventional beams are
designed to minimise the $\nu_e$ contamination in order to reduce the
background in the far detector.  Hence, the direct measurements of the
$\nu_e$ cross sections are based on samples that are 2-3 orders of
magnitude smaller than for $\nu_\mu$ and affected from larger
uncertainties coming from both the flux and detector response
\cite{ref82}, \cite{ref85}. The next generation of near detectors for
DUNE and Hyper-Kamiokande will make available larger samples so that
$\nu_e$ results will soon be systematic limited. At that time, the
impact of a high precision neutrino beam as \ENUBET will be
remarkable: it is summarized in Fig.~\ref{fig:nue}. Measurements for
electron antineutrinos are even more challenging, due to the lower
cross sections involved and the complete lack of measurements from
current experiments.  The \ENUBET negative polarity run will thus
provide the first precision measurement of $\sigma_{\bar{\nu}_e}$ with
a relative precision of $<5$\%.

\subsection{Search for new physics}

Disentangling models of new physics from the 3-flavour model
(i.e. sterile neutrinos, Non-Standard-Interactions - NSI -, unexpected
phenomena) requires mastering both the normalization and the spectrum
of the neutrino interaction rates~\cite{degouvea}. High precision
beams in the DUNE/Hyper-Kamiokande era will impact on this field both
indirectly and directly.  A high precision measurement of $\nu_e$
cross sections is mandatory in the presence of
NSI~\cite{ref85,ref87,ref83}, at long-baseline experiments.  Assuming
only standard interactions and exploiting cancellation of correlated
uncertainties between the $\nu_e$ and $\nu_\mu$ channels, the expected
number of electron neutrino interactions at far detectors can
currently be predicted with $\sim$ 5\% uncertainty~\cite{ref88} due to
the uncertainty of the $\sigma_e /\sigma_\mu$ ratio. A percent level
precision in the measurement of the $\sigma_e$ and $\sigma_\mu$ cross
sections will therefore enhance remarkably the sensitivity of
future long-baseline experiments to non standard effects.

In addition, \ENUBET will become the most urgent facility if the
Miniboone/LSND $\nu_e$ excess~\cite{Aguilar-Arevalo:2018gpe} is
confirmed by the Fermilab Short Baseline programme. \ENUBET provides a
complete control of the flavor at source and the NBOA technique gives
access to the oscillation pattern with a single detector in most of
the allowed Miniboone/LSND parameter space.

\subsection{Systematics reduction for the CP violation measurement}

A high precision neutrino source is a key asset for the measurement of
CP violation in the leptonic sector. CP violation is the core of the
physics programme at future long-baseline facilities (DUNE, Hyper-Kamiokande,
ESSnuSB) where the $\delta_{CP}$ accuracy relies on the possibility of
performing a precise determination of the rate of $\nu_e$ at far
detectors to pin down all the parameters of the $\nu_\mu\to\nu_e$ and
$\bar{\nu}_\mu\to\bar{\nu}_e$ oscillation probabilities.  The impact
of systematics in the CP discovery reach has been outlined in several
works~\cite{CPVsys0, CPVsys1}. Thanks to larger detectors and new or
upgraded beams the overall statistical error on appearance events will
finally reach the 1-2\% level \cite{Huber} thus becoming smaller than
the systematic uncertainty we can expect as of today.
  
The number of observed $\nu_e^{CC}$ candidate events at the
far detector can be expressed as:
\begin{equation}
  N_F(\nu_e) \propto \sum_{j}\left(\int \phi_F(\nu_\mu)P(\nu_\mu\to\nu_e) \sigma^{CC}_{j,\nu_e}\epsilon_j dE_{\nu} + B_j\right)
\end{equation}
where $j$ indicates the interaction mechanisms and all terms (flux
$\phi$, cross sections $\sigma$, efficiency $\epsilon$, oscillation
probability $P$) depend on the neutrino energy $E_\nu$ over which the
integral is performed.  The term $B$ is the background which, as well
as for genuine signal events, also depends on the knowledge of the
cross sections, the (oscillated) flux and the detector response.
Background is mostly due to the intrinsic $\nu_e$ component in the
initial flux and Neutral Current (NC) events with $\pi^0$ production.
The extraction of the physics encoded in the oscillation probability
term $P(\nu_\mu\to\nu_e)$ is thus intertwined with the
knowledge of cross sections and their energy dependence.

The near detectors
constrain the product of the unoscillated flux and the neutrino cross sections  (event rate at near
detector) and predict the rates at the
far detector.  At first approximation, the near-far detector comparison
cancels nuisance parameters that are not due to oscillation. The size
of second order corrections, however, is too large for the DUNE and HK
era. These corrections account for differences in the near and far
detector fluxes (due to finite distance effects), acceptances
(different detector and fiducial volumes), background, pile-up effects
due to the different neutrino rates and, in some cases, the different nuclear
average composition.
These factors can be mitigated building a ``far enough'' near detector
(intermediate detector) with the same detector technology of the far
detector. Still the composition of processes involved in the
background is different at the near and far site because the spectrum
of $\nu_\mu$ interactions is very different in the two locations due
to $\nu_\mu$ disappearance.
Far detectors are designed to maximize the statistics of signal
events. The designs of near detectors in long-baseline experiments aim
to provide much more detailed information on event kinematics,
which is mandatory to reduce uncertainties on fluxes, interaction
models and backgrounds. As a consequence, the near and far detectors
employ quite often different technologies \cite{ref9}, \cite{ref10}, \cite{ref72}.

The T2K Collaboration has been able to reduce the corresponding
uncertainties in the far detector rates down to 5-6\% in
the $\nu_\mu \to \nu_\mu$ disappearance channel and $\nu_\mu \to
\nu_e$ appearance channel \cite{ref88}.
The procedure of energy unfolding is quite involved: the link
between the observed event kinematics and the neutrino energy is
poorly known due to the lack of precise cross sections measurements.
For this reason, in most cases experiments provide flux integrated
cross sections as a function of variables that can be measured in a
straightforward manner, as the kinematical variables of the produced
lepton.  A workaround is offered by the above-mentioned NBOA technique
that provides a link between the (narrow) energy spectrum of the
neutrinos and the location of the neutrino interaction vertex at the detector.

In $\nu_\mu \to \nu_e$ appearance measurements, additional sources of
systematics arise from the difference in the final and initial
neutrino flavors. The near-to-far
event distribution ratio thus depends on the $\sigma_e$/$\sigma_\mu$ cross section ratio.
This ratio could be affected by the second-class currents or
deviations from the standard parameterization of the pseudoscalar
terms and, in general, it is not measured in a direct manner.

The \ENUBET approach consists in having a facility to measure the cross
sections as a function of energy with much better precision by
removing the flux normalization error and measuring a priori the energy spectrum
of neutrinos using an easy to reconstruct correlated variable
(radial distance of the vertex from the beam axis). This experimental strategy can be implemented
with a dedicated low intensity beam and a near detector employing the same detection technology as DUNE and HK. 

\section{$\nu_\mu$ cross section measurements}
\label{sec:numu}

Narrow-band beams are the ideal tool to perform high precision
$\nu_\mu$ scattering measurements since they provide a source where the
energy of the neutrino is known a priori with an uncertainty of
$\sim$10\%. The most striking drawback of narrow band beams is the
limited energy range covered by the neutrino source. Unfortunately,
the next generation of long baseline experiments requires a broad
energy coverage from 0.5~GeV to about 5~GeV. The most effective way to
increase the energy range spanning the entire region of interest for
DUNE and HK (0.5-5~GeV) is to employ the off-axis
technique~\cite{offaxis} in a single detector~\cite{Bhadra:2014oma} using a neutrino beam
where pions and kaons are selected by a transfer line with a narrow
momentum bite. This technique has been developed by \ENUBET in 2018 and
can be implemented using existing detectors as 
ProtoDUNE-SP and ProtoDUNE-DP at CERN.

\begin{figure}
\centering\includegraphics[width=0.8\textwidth]{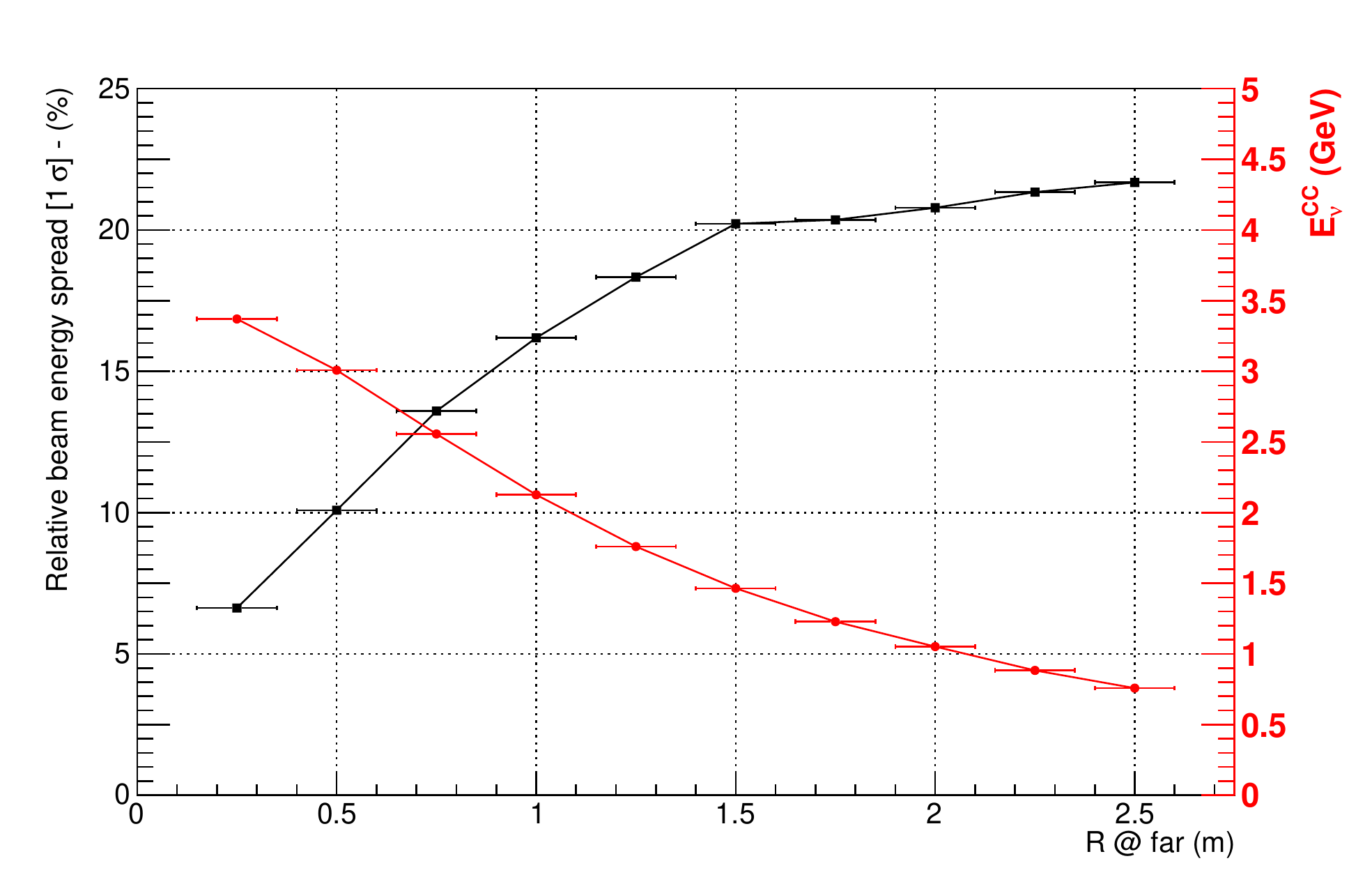} 
\caption{Beam energy spread (in black) and peak energy (in red) as a function of the distance $R$ of the interaction vertex at the detector from the
beam axis.}
\label{fig:en_res}
\end{figure}

The ``narrow band off-axis'' (NBOA)  technique~\cite{pupilli} exploits the strong
correlation between the energy of the neutrino interacting in the
detector and the radial distance ($R$) of the interaction vertex from
the beam axis in a 10\% momentum
bite beam.  The incoming neutrino energy is determined with a
precision given by the pion peak width of the spectrum at a fixed
$R$. In a transfer line optimized for the DUNE energy range (momentum of
secondaries: 8.5~GeV), it ranges from 7\% at 3.5 GeV to 22\% at 0.8
GeV, as illustrated in Fig.~\ref{fig:en_res}. Only a loose cut on the
visible energy is needed to separate the $\nu_\mu$ from pion decay
from the $\nu_\mu$ originating from the two body kaon decay. This cut
is not needed if the transfer line is optimized for the
Hyper-Kamiokande energy range (momentum of secondaries below 4 GeV)
since the kaon production is kinematically suppressed.  In this way, differential
cross section measurements can be performed without relying on the
reconstruction of the final state products for the determination of
the neutrino energy.  

The simplest implementation of this facility is based on a conventional
fast extraction (10~$\mu$s) horn over a narrow band transfer
line. This configuration can be implemented at CERN SPS (400~GeV
protons) and brings $77 \times 10^{-3} \ \pi^+$ per proton-on-target
(pot) for a single dipole, two triplet transfer line with a central
momentum of 8.5~GeV. This facility requires $1.1 \times 10^{19}$ pot
to produce $1.1 \times 10^6$ $\nu_\mu$ charged current (CC) events in
ProtoDUNE-SP. Due to the presence of the transfer line and the large
instantaneous currents produced during the fast extraction, the number
of secondaries reaching the decay tunnel can be measured by beam
current transformers with a precision of $\sim 1$\%.  Hence,
the measurement of the $\nu_\mu$ flux does not rely on simulation and
hadroproduction data except for the correction due to transported protons and kaons.

This configuration is compatible with current CERN North Area
infrastructures and with the size and position of ProtoDUNE-SP and
ProtoDUNE-DP in EHN1. Studies are ongoing to ascertain the overall
systematic budget on $\nu_\mu$ cross section measurements and, more
generally, the opportunities offered by this facility on short
baseline neutrino physics~\cite{near_detector}.

\section{$\nu_e$ cross section measurements}

A monitored neutrino beam is the ideal tool to measure neutrino cross
sections at percent level precision. It combines the above-mentioned
NBOA technique with a direct measurement of the $\nu_e$ flux and
provides a fully controlled source of $\nu_e$ at the GeV scale.

\begin{figure}
\centering\includegraphics[width=\textwidth]{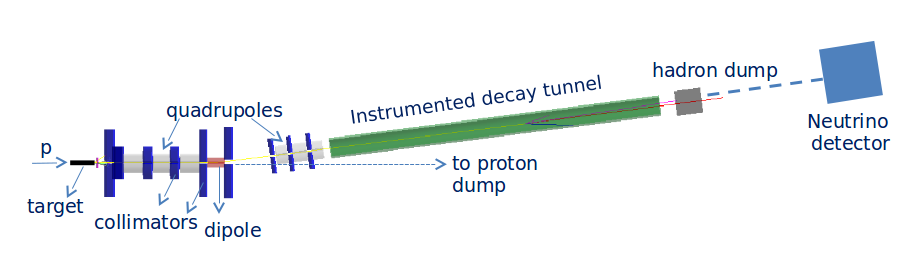} 
\caption{Schematics of the \ENUBET neutrino beam in the static focusing option (not to scale).}
\label{fig:beamline}
\end{figure}

The \ENUBET neutrino beam (see Fig.~\ref{fig:beamline}) is a
conventional narrow band beam with a short ($\sim$20~m) transfer line
followed by a 40~m long decay tunnel. Unlike most of the beams
currently in operation, the decay tunnel is not located in front of
the focusing system (horns) and the proton extraction length is slow:
a few ms in the horn option and 2~s in the static focusing option (see
below). Particles produced by the interaction of protons on the target
are focused, momentum selected and transported at the entrance of the
tunnel. Non-interacting protons are stopped on a proton beam dump.

The particles that reach the decay tunnel are hence pions, kaons and
protons within the momentum bite of the transfer line (10\% in
ENUBET). Off-momentum particles are mostly low energy pions, electrons,
positrons and photons from tertiary interactions in the collimators and
other components of the beamline, and muons from pion decay that cross
the collimators. Due to the presence of the transfer line and the long
proton extraction time ($>2$~ms), the rate of these particles are
several orders of magnitude smaller than beams currently in operation
and the instrumentation located in the decay tunnel can monitor lepton production at single
particle level.

Kaon decays are particularly well suited for single-particle
monitoring.  In the current \ENUBET design (tuned for the energy range
of interest of DUNE) the mean energy of the hadrons selected in the
transfer line (8.5~GeV) and the length of the decay tunnel is
optimized in order to have only one source of electron neutrinos: the
$K_{e3}$ decay of the kaons - $K^+ \rightarrow \pi^0 e^+
\nu_e$. Electron neutrinos from the decay in flight of kaons represent
$\sim 97$\% of the overall $\nu_e$ flux. Since the positrons are
emitted at large angles with respect to muons from pion decay,
particles produced by the kaons reach the wall of the instrumented
tunnel before hitting the hadron dump (see
Fig.~\ref{fig:beamline}). The vast majority of undecayed pions,
particles transported along the transfer line and muons from $\pi^+
\rightarrow \mu^+ \nu_\mu$ reach the hadron dump without hitting the
walls and do not contribute to the particle rate in the instrumented
walls.

The rate of positrons from $K_{e3}$ decays is monitored at single
particle level by longitudinally segmented calorimeters that separate
positrons from pions, muons, neutrons and protons. The modules of the
calorimeter are located inside the beam pipe and are assembled into
cylindrical layers. Positron/photon separation is performed by a
photon veto made of plastic scintillator tiles located just below the
innermost layer.

Particle rates in the instrumented walls are sustainable only if the
proton extraction is $\gg 10$~$\mu$s. \ENUBET is considering two
focusing options. The first option is based on a magnetic horn that is
pulsed for 2-10 ms and cycled at several Hz during the accelerator
flat-top. This option has a very large acceptance producing at SPS $77 
\times 10^{-3} \ \pi^+$/pot and $7.9 \times 10^{-3} \ K^+$/pot. The proton
extraction scheme (``burst mode extraction'') has been studied at the
CERN-SPS in 2018 and will be commissioned after the LHC Long
Shutdown 2. In summer 2018, \ENUBET also demonstrated the effectiveness
of a purely static focusing system based on DC operated magnets. The
static option allows for slow proton extractions (2-4 s) reducing by
two order of magnitude the rate at the instrumented decay tunnel.
Static focusing is the ideal focusing scheme for any monitored
neutrino beam and paves the way (see below) to the first tagged
neutrino beam. It has, however, a smaller acceptance than the
horn-based system producing $19 \times 10^{-3} \ \pi^+$/pot and $1.4
\times 10^{-3} \ K^+$/pot. As a consequence, it requires about $4.5
\times 10^{19}$ pot at the SPS to carry out the cross section program
both with $\nu_e$ and $\nu_\mu$. Similarly, the cosmic ray veto system
of the ProtoDUNE detectors should be improved to suppress cosmic ray
background without relying on beam timing information.

In \ENUBETnospace , the rate of positrons provides a direct measurement of the $\nu_e$
produced in the tunnel. The distribution of particles (positron,
muons, pions) along the axis of the tunnel and their energy and polar
angle distribution constrain any source of systematic bias between the
rate of positrons observed in the tunnel and the expected rate of
$\nu_e$ at the detector. Unlike present beams, at leading order no
information is needed from particle production yields
 in the target (hadro-production), the simulation of transport and
reinteraction of secondary particles in the beamline, the monitoring
of the protons on target and of the currents in the horn because the
rate of particle production in the tunnel provides an observable that
is directly linked to the flux. \ENUBET is now studying sub-leading
effects to demonstrate that the total systematic budget of the $\nu_e$
flux is below 1\%. 

Since the $K_{e3}$ branching ratio is known with a precision of 0.8\%,
positron monitoring also provides the total production rate of kaons
at the per-cent level. This precision can be further improved
monitoring the rate of pion production in the tunnel due to the other
decay modes of kaons and, in particular, the leading $K^+ \rightarrow
\mu^+ \nu_\mu$ (BR~$\simeq$~63\%) and $K^+ \rightarrow \pi^+ \pi^0$
(BR~$\simeq$~21\%). These channels, which were not included in the
original physics programme of \ENUBETcomma~are now exploited to evaluate
the flux of $\nu_\mu$ from kaons.  Finally, a direct measurement of
the rate of muons from $\pi^+$ decays after the hadron dump cannot be
done at single particle level for the horn-based option but it can be
performed if the focusing is purely static because the muon rate is
reduced by two order of magnitudes. In this case, the muon rate after
the hadron dump provides the $\nu_\mu$ flux from pion decays with a
precision comparable with the $\nu_e$ flux.

\ENUBET can be operated in inverse-polarity mode, producing
$\bar{\nu}_e$ from $K^- \rightarrow \pi^0 e^- \bar{\nu}_e$ and
$\bar{\nu}_\mu$ from $\pi^- \rightarrow \mu^- \bar{\nu}_\mu$ and $K^-
\rightarrow \mu^-\bar{\nu}_\mu$. All considerations above apply to
\ENUBET as a source of electron and muon anti-neutrinos.
 
\section{Tagged neutrino beams}

A purely static focusing system opens up several opportunities beyond
the original goals of \ENUBETnospace . Since the proton extraction can be diluted up to several seconds, the instantaneous rate of large angle
decay products in the decay tunnel is reduced by about two orders of
magnitude compared with the horn option. In the \ENUBET static option
the time between two $K_{e3}$ decays is 1.3~ns, which can be further
increased operating with a 4~s extraction or a smaller number of pot
per cycle. The occurrence of a neutrino interaction in the detector can thus be time linked with the observation of the lepton in the decay
tunnel. Such an observation has never been performed in any neutrino
experiment at any energy and would represent a major breakthrough in
experimental neutrino physics. A facility where the neutrino is
uniquely associated with the other decay particles of the parent kaon
is called a ``tagged neutrino beam''. Physicists have speculated
about this possibility just after the first direct observation of
neutrinos~\cite{hand1969,Pontecorvo:1979zh} but the technologies that
provide time resolution, pile-up mitigation and radiation hardness for
time tagged neutrino beams are available since a few years only. In
order to suppress accidental coincidences between the neutrino
interaction and uncorrelated particles inside the beam pipe, the
timing precision of the detectors that are used to instrument the
decay tunnel must reach 100~ps. This precision is
needed also to associate the positron with the other decay products of
the kaons.  The timing precision of the
neutrino detector should not exceed a few ns, as well. The physics potential of tagged neutrino beams are
outstanding since they provide energy and flavor measurement on an
event-by-event basis and are the ideal tool to study cross sections
and non standard oscillation phenomena, including sterile
neutrinos. For the first time,  they also give experimental access to the lepton-neutrino
entangled state to study propagation and collapse of the
wavefunction. A CERN based implementation requires an improvement of
the timing resolution of the ProtoDUNE Photon Collection System up to $\sim$~1~ns and a significant 
R\&D effort since this technology is not mature as the monitored neutrino beams.

\section{Conclusions}
\label{sec:conclusion}

Neutrino oscillation physics gathers one of the largest community in
particle physics and CERN is playing a leading role in the
construction of next generation long baseline neutrino experiments.  The physics reach of these experiments can be substantially enhanced by
percent level precision measurements of the $\nu_\mu$ and $\nu_e$
cross sections. These precisions are mandatory to extract oscillation
parameters and non-standard effects from the rate of $\nu_e$ and
$\nu_\mu$ at the far detector of DUNE and Hyper-Kamiokande. They also
ground on solid base our understanding of electroweak nuclear physics
and the forthcoming results on CP violation in the leptonic sector. A
dedicated short baseline facility for a new generation of cross
section experiments is therefore the most cost-effective completion of
the long-baseline programme.  A high precision narrow band beam with
lepton monitoring at single particle level is the ideal tool for
systematic reduction in the DUNE and Hyper-Kamiokande era and will
bring to a major leap in our knowledge of neutrino cross
sections. CERN is the natural candidate to host such facility at SPS
given the outstanding particle identification capability, energy
resolution and fiducial mass of the liquid argon detectors operated in
EHN1.

\section{Acknowledgements}

The ENUBET project has received funding from the
European Research Council (ERC) under the European Union's Horizon
2020 research and innovation programme (grant agreement N. 681647).

\end{document}